**Early formation and recent starburst activity in the nuclear disc of the Milky Way.**

F. Nogueras-Lara[1], R. Schödel[1], A. T. Gallego-Calvente[1], E. Gallego-Cano[1], B. Shahzamanian[1], H. Dong[1], N. Neumayer[2], M. Hilker[3], F. Najarro[4], S. Nishiyama[5], A. Feldmeier-Krause[6], J. H. V. Girard[7] and S. Cassisi[8]

[1]Instituto de Astrofísica de Andalucía (CSIC), Glorieta de la Astronomía s/n, 18008 Granada, Spain
email: nogueras@mpia.de
[2]Max-Planck Institute for Astronomy, Königstuhl 17, 69117 Heidelberg, Germany
[3]European Southern Observatory (ESO), Karl-Schwazschild-Strasse 2, 85748 Garching, Germany
[4]Departamento de Astrofísica, Centro de Astrobiología (CSIC-INTA), Crta. Torrejón a Ajalvir km 4, 28850, Torrejón de Ardoz, Spain
[5]Miyagi University of Education, Aoba-ku, 980-0845 Sendai, Japan
[6]The University of Chicago, The Department of Astronomy and Astrophysics, 5640 S. Ellis Ave, Chicago, IL 60637, USA
[7]Space Telescope Science Institute, Baltimore, MD 21218, USA
[8]INAF-Astronomical Observatory of Abruzzo, Via M. Maggini, sn.64100 Teramo, Italy

**The nuclear disc is a dense stellar structure at the centre of the Milky Way, with a radius of ~150 pc[1]. It has been a place of intense star formation in the past several tens of millions of years[1,2,3] but its overall formation history has remained unknown up to now[2]. Here we report the first detailed star formation history of this region. The bulk of its stars formed at least eight billion years ago. This initial activity was followed by a long period of quiescence that was ended by an outstanding event about 1 Gyr ago, during which roughly 5% of its mass formed in a time window ~100 Myr, in what may arguably have been one of the most energetic events in the history of the Milky Way. Star formation continued subsequently on a lower level, creating a few percent of the stellar mass in the past ~500 Myr, with an increased rate up to ~30 Myr ago. Our findings contradict the previously accepted paradigm of quasi-continuous star formation at the centre of the Milky Way[4]. The long quiescent phase agrees with the overall quiescent history of the Milky Way[2,5] and suggests that our Galaxy's bar may not have existed until recently, or that gas transport through the bar was extremely inefficient during a long stretch of the Milky Way's life, and that the central black hole may have acquired most of its mass already in the early days of the Milky Way.**

The GALACTICNUCLEUS survey[6] has been specifically designed to study the structure and formation history of the Galactic centre (GC) with sensitive, high angular resolution (0.2" full width at half maximum), near-infrared images. Here we analyse a rectangular region of ~ 37.5′ × 8.3′ (about 1600 pc$^2$ at the distance of the GC) centred on Sagittarius A* (Sgr A*, 17$h$ 45$m$ 40.05$s$, -29◦ 00′ 27.9", Fig. 1, bottom), the Milky Way's central black hole[7,8,9]. A region close to Sgr A* was excluded from the analysis because of the extreme stellar crowding, so were regions dominated by dark foreground clouds. The final catalogue includes accurate photometry for more than 700,000 stars, superseding other surveys of the GC by orders of magnitude[6,10].

The $HKs$ colour magnitude diagram (CMD, Fig. 1, top left) shows a prominent vertical feature at $H$-$Ks$ > 1.3, which corresponds to giant stars located at the GC. Stars with bluer colours lie in spiral arms in the foreground[6,10]. The diagonal high-density feature that extends from roughly ($Ks$=14.5, $H$-$Ks$ =1.3) to ($Ks$ = 16.5, $H$-$Ks$ =3) is caused by differential extinction and reddening of the so-called red clump (RC), which marks the location of helium core burning giants in metal-rich, old populations. Stars in the RC can be used as standard candles because of their highly similar luminosities and temperatures across a broad range of ages, metallicities, and masses[11]. We used these stars to correct our data for the effects of interstellar extinction and reddening, and finally created a so-called $Ks$-band luminosity function (KLF, corrected for completeness, Fig. 1, top right), which reflects the number of stars in given brightness bins.

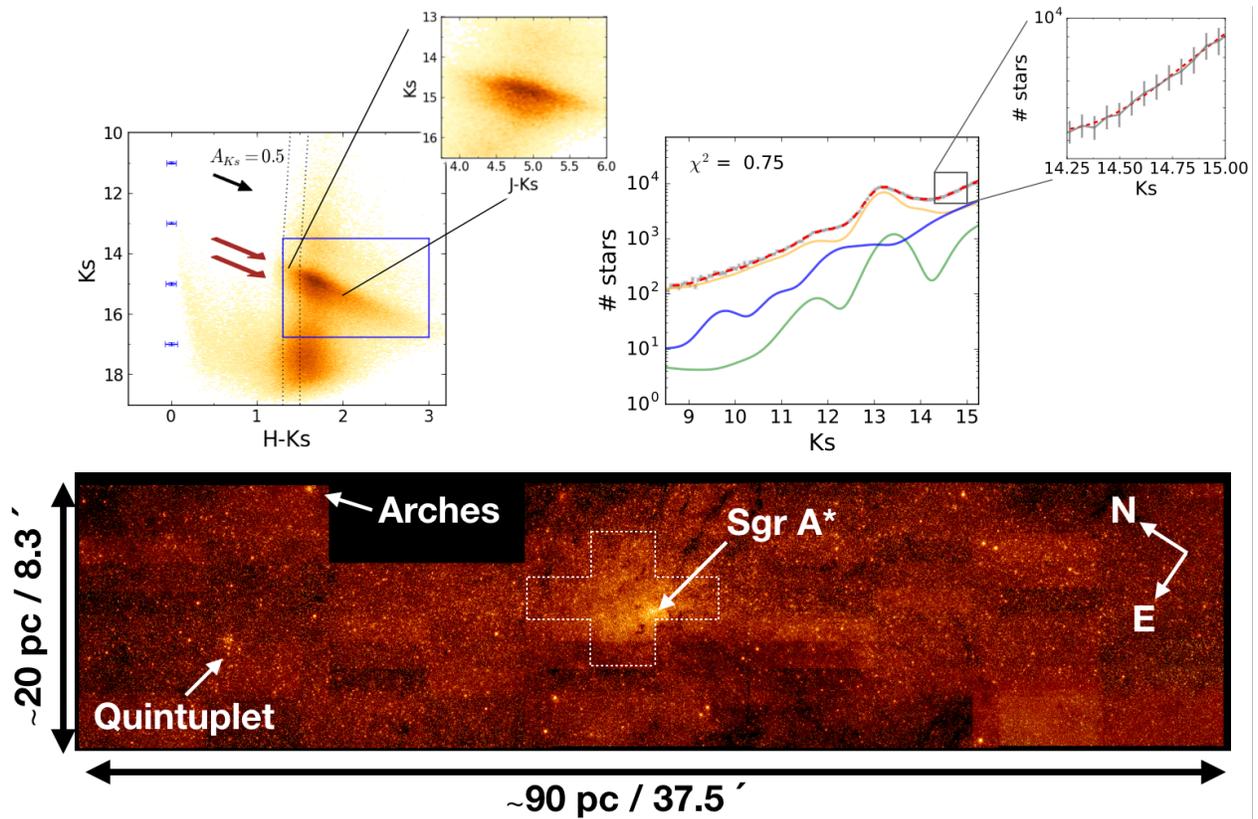

*Fig. 1: Lower panel: Ks-band image of the GALACTICNUCLEUS survey region examined for this study. The positions of Sgr A*, the Arches and the Quintuplet clusters are indicated. The cross-shaped region outlined by a white dashed line was excluded because of too high source crowding. The dark rectangle right next to the Arches cluster is a field with incomplete data. Upper left: Observed H-Ks CMD. The vertical dashed lines show the two colour cuts used to exclude foreground stars (see Methods). The two arrows indicate the direction of reddening and point toward the two parallel features that we detected in the RC, corresponding to old (brighter) and younger (fainter) stars. The inset shows the double RC in the J-Ks CMD. The blue rectangle indicates the stars used to create the extinction map. The error bars, in blue, show the uncertainties for different magnitude bins. Upper right: KLF (grey), obtained after extinction correction of the observed stars, and best-fit model of the star formation history (red dashed line). For simplicity we combined the theoretical LFs used into old, intermediate and young ages (orange, green and blue, respectively). The zoom shows the quality of the data in detail, including error bars and model fit.*

The luminosity function of a stellar population changes depending on its age. Therefore, the observed KLF can be interpreted as the superposition of the LFs from star formation events at different times. It thus contains the imprint of the formation history of the nuclear disc. We fitted linear combinations of theoretical single age LFs to our measurements. To deal with the systematic uncertainties related to the models, we used stellar evolutionary models from two different groups (BaSTI[12] and MIST[13,14,15,16,17]), which resulted in very similar histories, but with somewhat different ages for specific events. Our model fits require a mean metallicity of the nuclear bulge stars of about twice solar, in agreement with what has been found for the innermost Galactic bulge[10].

The derived star formation history (SFH) is summarized in Fig. 2. Over 80% of the stars formed between 8 – 13.5 Gyr ago. This initial formation was followed by a quiescent period of more than 6 Gyr duration, which was ended by a remarkable event 600 Myr or 1 Gyr ago (depending on the models used), during which ~5% of the stellar mass formed (in the following we refer to this as the "1 Gyr event"). Star formation continued in the recent past, but on a lower level. The past 30 Myr have been relatively active with a star formation rate 0.2-0.8 $M_\odot$/yr (using BaSTI and MIST models to estimate the uncertainties). This is in good agreement with constraints from classical Cepheids[3], and requires that a few $10^6$ $M_\odot$ stars formed in the recent past in the GC. This implies that, in addition to the known young, massive clusters (the Arches, Quintuplet, and Central clusters), there must be a large number of clusters that have eluded detection so far because of their partial dissolution in a heavily crowded field[18].

The 1 Gyr event has left a visible imprint in the CMD (Fig. 1, upper left panel). The RC sequence is double (see Methods section), with a weaker one, corresponding to stars created in the 1 Gyr event, running parallel to the primary one, about 0.5 magnitudes fainter, that corresponds to the oldest stars.

The nuclear disc region within 120 pc of Sgr A* contains $(8\pm2)\times10^8$ $M_\odot$ based on near-infrared flux densities[1], which we confirm here using independent, mid-infrared measurements (Methods section). If we scale the ~1600 $pc^2$ studied here (Fig. 1) to the ~17000 $pc^2$ corresponding to the 120 pc radius, then our analysis of the KLF results in a fully consistent mass of $(9.8\pm0.6)\times10^8$ $M_\odot$. If the 1 Gyr event comprised the same area, then it created stars with a total mass of $(4.2\pm0.7)\times10^7$ $M_\odot$. Our modeling shows that the duration of this event was probably shorter than 100 Myr. The star formation rate was thus 0.4-0.5 $M_\odot$/yr, not much lower than the ~1-2 $M_\odot$/yr assumed for the entire Galaxy at present time[2], but limited to less than 1% of its volume. Hence, scaled to the size of the Milky Way, the conditions in the GC during the 1 Gyr event must have resembled those in starburst galaxies, that form stars at rates ≳ 100 $M_\odot$/yr. As a consequence, the 1 Gyr event must have resulted in ≳$2\times10^5$ core collapse supernovae and thus have driven a powerful outflow from the GC region.

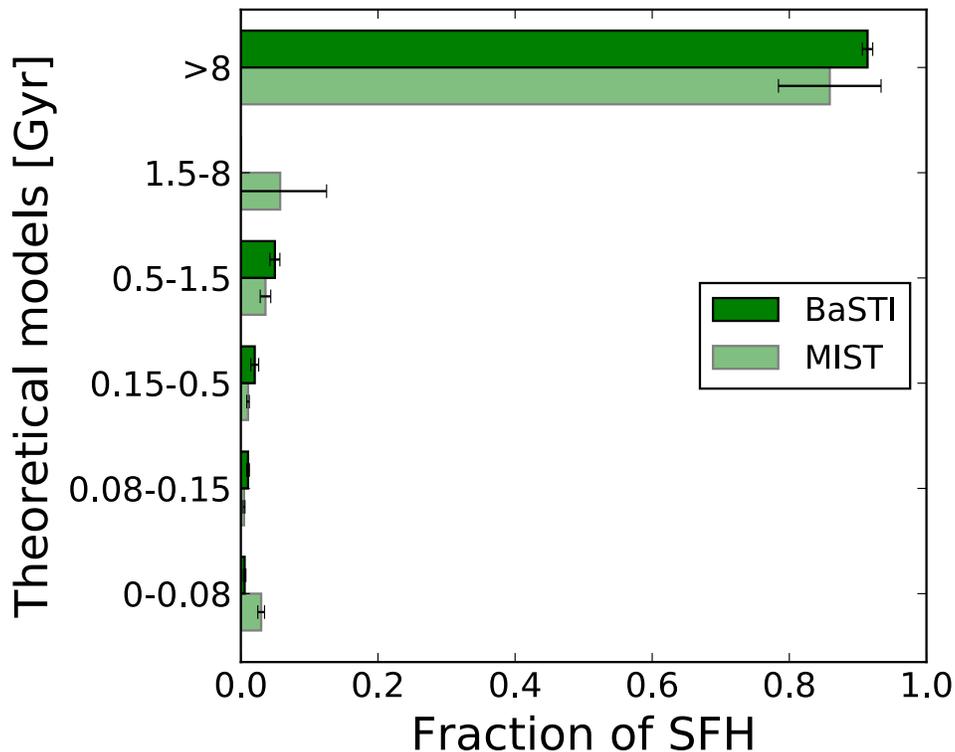

Fig. 2: Star formation history of the nuclear stellar disc as derived from the model fits to the KLF with BaSTI and MIST isochrones. There was no contribution for any of the BaSTI models in the range 1.5-8 Gyr.

Our results show that the nuclear disc formed very early in the Galaxy's history, and may have a similar age than the much larger Galactic bulge[19], in which it is embedded. The following quiescent period suggests that gas transport toward the GC was suppressed or gas was expulsed efficiently. Even in the absence of direct gas inflow, gas from stellar winds is expected to accumulate at the bottom of the potential well of the Milky Way. Occasional activity of Sgr A*, may have played a role in driving out large quantities of gas via an accretion-powered central wind[20]. However, the bar of the Milky Way is thought to be very efficient at transporting gas toward the GC, at a rate of 0.1-1 $M_\odot$/yr[21] and it is therefore questionable whether black hole activity alone could have suppressed star formation in the nuclear disc during billions of years. Our findings therefore suggest that gas transport through the Galactic bar to the GC was inefficient during billions of years, which may have a variety of reasons, which range from stalling of gas transport followed by star formation near resonances to the non-existence of the bar. A recent analysis of Gaia DR2 data indicates that the Galactic bar may be currently still forming or has been strongly disturbed recently, possibly by the Sgr dwarf satellite galaxy or a similar perturber[22,23].

The Milky Way has not suffered any major merger (20% of its mass or more) over the past ~10 Gyr[5,24], but minor interactions may have been more frequent. In this context it is relevant that simulations of the orbit of the Sgr dwarf galaxy indicate that it may have reached minimum separation from the GC roughly 1 Gyr ago[25]. This event, or a similar one, may have triggered gas infall into the GC, followed by star formation.

Radio, X-ray, and γ-ray observations have found the signatures of strong outflows from the Galactic Centre region[26]. The so-called Fermi Bubbles, for example, are features traced in microwaves, X-rays, and γ-rays that extend to about 10 kpc on both sides of the Galaxy and emanate from its centre[27,28]. At their bases they appear to connect to X-ray structures, which have sizes corresponding to the scale of the nuclear disc and may trace exhaust channels of hot plasma[26]. Our results can put these observations into context. One explanation for the origin of the Fermi Bubbles was based on the previous paradigm of quasi-continuous star formation at the GC, which would create the Fermi Bubbles through a continuous injection of cosmic ray protons[29]. However, this hypothesis requires a quasi-continuous star formation at the GC sustained over more than 5 Gyr and can therefore be ruled out because its basic assumption contradicts the SFH derived here. On the other hand, we estimate that ~$3\times10^4$ supernovae exploded in the GC in the past 30 Myr, with an energy input of ~$10^{51}$ erg each, on average. A small fraction of this energy will be sufficient to account for the energy in the X-ray outflows. Therefore, recent star formation, possibly combined with some activity of Sgr A*, can provide a plausible explanation for the observed outflows[30].

We find that ≳90% of the stellar mass of the GC were in place 8 Gyr ago. Since black holes grow mostly through the accretion of gas and star formation requires the presence of gas, the low star formation rate over the following Gyr may imply that the central black hole, Sgr A*, had already acquired most of its present day mass in the early days of our Galaxy.

This is the first time that we have gained some detailed insight into the formation history of the nuclear disc in the GC. Future spectroscopic and high angular resolution imaging follow-up observations will be able to constrain the different events further and teach us about the formation history of the GC and its implications on the evolution of the Milky Way and its supermassive black hole.

**Data Availability**

All the raw data used in this study are available at the ESO Science Archive Facility (http://archive.eso.org/eso/eso_archive_main.html) under programmes IDs 195.B-0283 and 091.B-0418. The final version of the GALACTICNUCLEUS survey (images and point source catalogues) will be released to the public via the ESO Phase 3 platform within the next year.

**Author contributions**

F. N.L. Reduced the data and produced the catalogue, carried out the main part of the analysis, and wrote the draft version of the manuscript. R. S. planned the research project, collaborated in the data analysis and interpretation, and organised the writing of the manuscript. A. T. G.-C. collaborated in the data reduction. S. C. Computed the theoretical LFs. E. G-C., B. S., H. D., N. N., M. H., F. N., S. N., A. F.-K. and J. H. V. G. participated actively in the scientific discussion and interpretation and contributed to the production of the final version of the manuscript.

**Acknowledgements**

This work has made use of BaSTI web tools. The research leading to these results has received funding from the European Research Council under the European Union's Seventh Framework Programme (FP7/2007-2013) / ERC grant agreement nº [614922]. This work is based on observations made with ESO Telescopes at the La Silla Paranal Observatory under programmes IDs 195.B-0283 and 091.B-0418. We thank the staff of ESO for their great efforts and helpfulness. F. N.-L. acknowledges financial support from a MECD pre-doctoral contract, code FPU14/01700. We acknowledge support by the State Agency for Research of the Spanish MCIU through the "Center of Excellence Severo Ochoa" award for the Instituto de Astrofísica de Andalucía (CSIC) (SEV-2017-0709).

**METHODS**

**I Data**

For this work we have used the *J-*, *H-* and *Ks*-band photometry of 14 different fields from the GALACTICNUCLEUS survey[6]. This is a near infrared, high angular resolution (~0.2'') survey of the Galactic Centre carried out with the High Acuity Wide-field K-band Imager located at the Very Large Telescope (HAWK-I/VLT). The speckle holography algorithm is key for this survey to overcome atmospheric seeing. The high angular resolution of the GALACTICNUCLEUS data reduces crowding by a factor of ≳ 10 compared to seeing limited surveys. As a consequence, GALACTICNUCLEUS reaches several magnitudes deeper than other surveys of the same region[6,10]. All the data will be publicly available via the ESO science archive facility within the next year. The photometric uncertainties are less than 0.05 mag at $J\sim20$, $H\sim17$ and $Ks\sim16$, with a zero point uncertainty of 0.036 mag in all three bands. We have produced the final photometric catalogue by combining all fields, correcting for possible photometric offsets between them.

**II Removing dark clouds**

We excluded regions with dark clouds because they are largely opaque and limit the observation of stars belonging to the nuclear stellar disc (NSD). Given that the extinction is larger for shorter wavelengths, we used a *J*-band density map to identify dark clouds[6]. We defined a pixel size of ~6'' and compute the *J*-band density for each pixel. We masked regions with less than 5% of the maximum density detected (Fig. 3). Varying this percentage between 3% and 7% did not result in any significant change to our results.

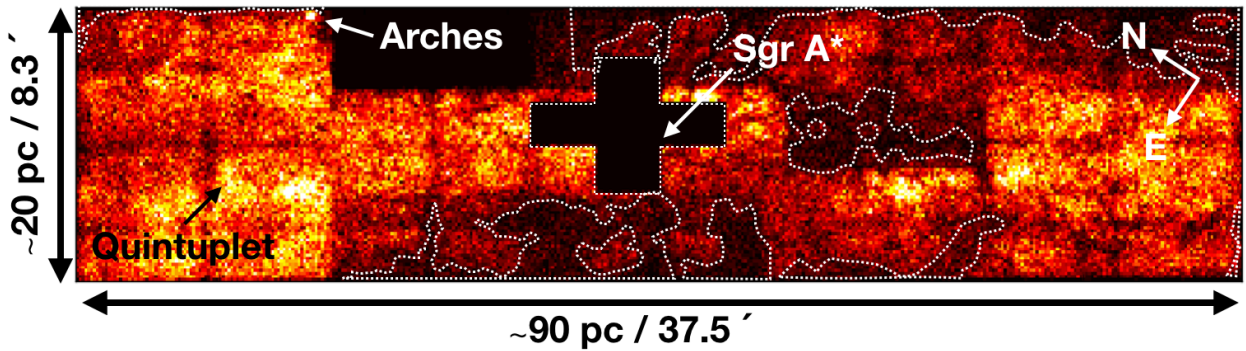

*Fig. 3: J-band density map. The dark patches are clouds of molecular gas and dust. They impede the view into the GC and were therefore excluded from this study (white dashed line contours). The central cross-shaped region corresponds to a low completeness region.*

**III Extinction map**

As described in our previous work, we used RC stars to compute an extinction map to de-redden the photometry and to correct for differential extinction[6, 10]. We selected a pixel scale of 2.5''pixel$^{-1}$ and computed the colour excess H-Ks using only stars within 5'' from a given pixel. Since the differential extinction and the extinction in the line of sight can vary significantly on arc second scales, we considered only stars whose H-Ks value lies within 0.25 mag of the value corresponding to the star closest to any given pixel. We imposed a minimum of five stars fulfilling these conditions per pixel on the extinction map. We weighted the distances to the centre of the pixel with an inverse-distance weight (IDW) method[6, 10]. We used the RC stars shown in the rectangle in Fig. 1 (upper left) and assumed an extinction index of 2.30±0.08[6]. This region is mainly populated by RC stars or RGBB stars whose intrinsic colours are similar[10]. We computed the intrinsic colour (H-Ks)$_0$= 0.07 ± 0.03 taking into account different metallicities and stellar ages of RC and RGBB stars, BaSTI models and the corresponding filters of HAWK-I. Fig. 4 shows the obtained extinction map. The statistical uncertainties of the pixels are ~3% and were calculated taking into account the dispersion of the stars used to compute the extinction value of each pixel and the standard deviation of the intrinsic colour computed previously. The systematics are ~6% and considers the systematic uncertainty of the zero point for H and Ks, and the uncertainties of the extinction index and the effective wavelength. These uncertainties do not influence the differential extinction in any significant way.

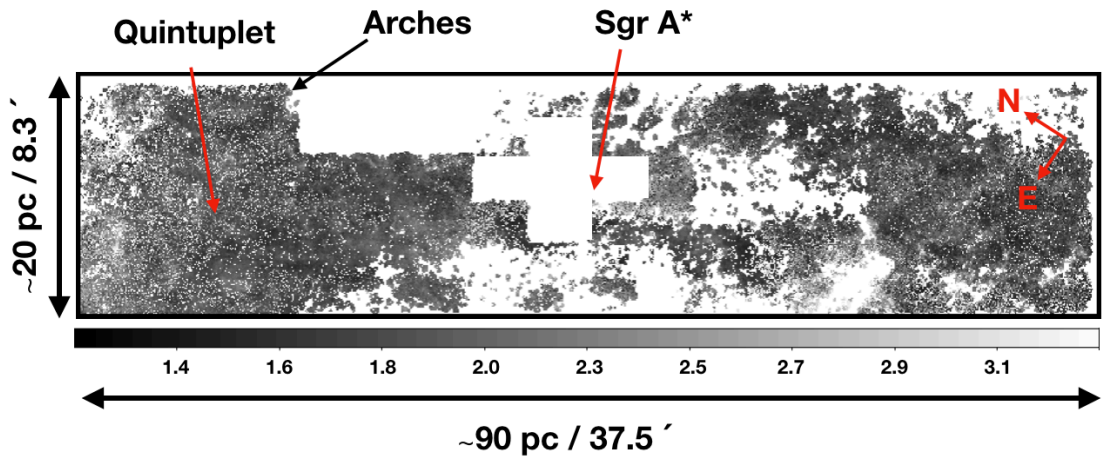

*Fig. 4: Extinction map $A_{Ks}$. White patches and pixels indicate regions where the number of stars was not enough to compute an extinction value. Those regions mainly correspond to dark clouds and were previously excluded from our analysis.*

After applying the extinction map to the CMD, we obtained the de-reddened CMD shown in Fig. 5.

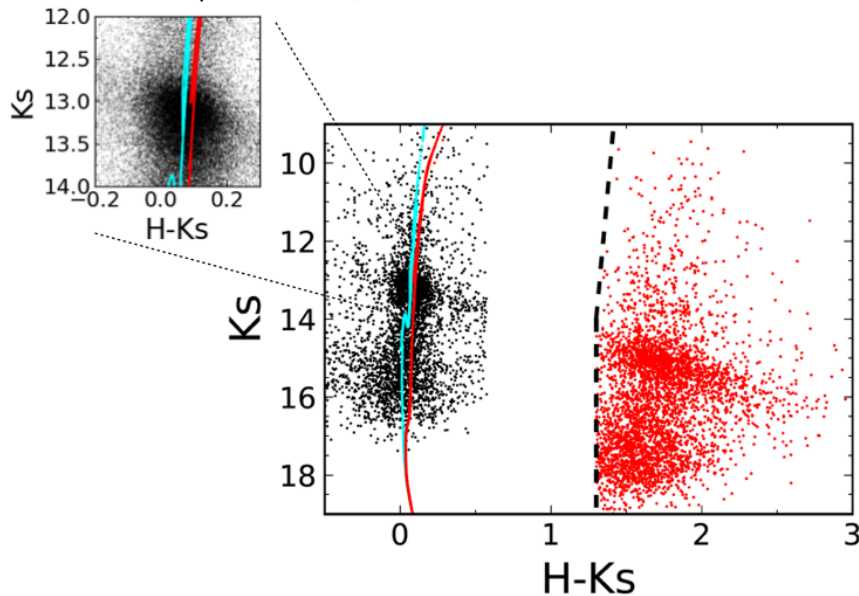

*Fig. 5: CMD Ks versus H-Ks showing original data (red dots) and the result after applying the extinction correction (in black). The black dashed line shows the cut in H-Ks to exclude the foreground population.*

*The red and cyan lines correspond to BaSTI isochrones of an old and a young population ~10 Gyr and ~1 Gyr, respectively according to our results. Only a random fraction of the total amount of stars is shown for clarity.*

**IV Luminosity Function**

*Masking of dark clouds*

To obtain the KLF, we masked the region closest to Sgr A* (the nuclear star cluster), where the extreme source crowding limits the completeness of our data significantly at magnitudes Ks ≳ 15.

*Foreground stars*

About 50% of the interstellar extinction toward the GC is caused by dust in the Galactic Disc. The bar/bulge is relatively dust free and roughly the other 50% of extinction is caused by dust in the Central Molecular Zone, which forms already part of the GC[1]. The HKs CMD (Fig. 1) shows how stars in spiral arms in the foreground Galactic Disk appear only in significant numbers at *H-Ks* ≲ 1.

*We have chosen a colour cut of H-Ks = 1.3 to exclude foreground stars. The most extreme intrinsic colours in H-Ks, about 0.2-0.3 mag, are reached M-type dwarfs or giants. In order to contaminate our sample, those stars would have to suffer a reddening of H-Ks > 1 mag and thus lie in the innermost parts of the Disk or outer layers of the bar. However, M-type main sequence stars would be many magnitudes fainter than our detection limit and can thus be ruled out as contaminants. M-type giants are very bright and rare. We would expect to see them in the top of the H-Ks CMD. However, there are only stars with H-Ks > 1.5 in this part of the CMD. With our chosen colour cut we can therefore safely exclude contamination by foreground stars in the Galactic Disk.*

What about contamination by stars in the foreground bar/bulge? We have shown in previous work that the extinction toward the innermost parts of the bar (at ~90 pc vertical separation from Sgr A*) is $A_{Ks}$ ~1.2, almost half the mean extinction as measured for the GC field studied here[10]. With the reddening curve inferred for this work $A_{Ks}$=1.2 corresponds to H-Ks = 1.0. With the scatter from measurement uncertainties, stars from the innermost bulge/bar may therefore contaminate our sample to a certain degree. We consider that this potential contamination does not pose any serious problem for our analysis because those contaminating stars would belong to a ~12 Gyr old single age, metal-rich stellar population[10] and would thus only affect the oldest age bin in the SFH. If we could cleanly subtract the contaminant stars, we would at most expect a decrease of the number of or a tendency toward somewhat young ages stars in the oldest age bin. We tested this effect by choosing a stricter colour cut of H-Ks = 1.5 and repeating our analysis of the SFH. Indeed, the age of the oldest stars tend to be somewhat younger (moving from ~13 Gyr to 11 Gyr), but the rest of the SFH is unaffected.

In addition, to obtain a rough estimate of the potential contamination by foreground stars, we varied the colour cut by Δ(H-Ks) = 0.2 mag. This corresponds to the standard deviation, $\Delta A_{Ks}$ = 0.25 mag, of the mean extinction determined for our survey region (see above). The number of stars that get included into or excluded from the final sample varies then by about 2%. We generated and analysed the KLFs for both cases and applied the KLF fitting with MC estimation of uncertainties as described above. We did not find any significant difference with respect to our results.

For creating the KLF, we use all stars detected at *Ks* and do not require that they are detected at *H*, too. Since reddening rises steeply toward shorter wavelengths, highly reddened stars may not have any *H-* band counterparts. We can safely assume that stars without any *H*-band counterpart are not in the foreground.

Finally, it is important to point out that we are dealing with a complex situation here: The bar/bulge and the NSD cannot be spatially separate features and stars from the bar are expected to be present throughout the NSD. As described above this will not change any of our conclusions in a significant way because the contaminating population is old, single age, and metal rich.

*Saturation*

We corrected for saturation in the *Ks*-band by using the SIRIUS/IRSF survey[31] to replace the HAWK-I photometry for bright stars (*Ks* < 11.5 mag).

*Completeness*

Given the large size of the images, the great number of stars, the variability of stellar density with position, and the variation of the data quality for each field and filter, we considered the standard method of determining completeness through inserting and recovering of artificial stars as unviable because of its complexity and the enormous amounts of computation time necessary for such a test (we estimate several weeks on our multi core server). We therefore used a previously developed alternative approach, which is based on determining the critical distance at which a star of any given magnitude can be detected around a brighter star[32]. This critical distance can then be translated into completeness maps for each magnitude bin analysed. This method assumes that the probability of detecting a source with a given magnitude is uniform across the analysed field. Since the stellar density is non-uniform across our survey area and the quality of the data depends on the specific observing conditions and filter for each of the 14 fields that make up the final mosaic, we divided the entire field into 108 smaller sub-regions of 2'x1.4'. We considered these sub-regions sufficiently small to assure the validity of the basic assumption of constant stellar density and homogeneous stellar population. To estimate the overall completeness of our star counts and its related uncertainty, we then averaged over the completeness solutions obtained for each of the sub-regions and used the corresponding standard deviation as uncertainty (Fig. 6).
To test the reliability of our approach, we performed two independent tests in the central field of the mosaic, where crowding is most significant: (1) We estimated completeness through a standard artificial star test and found that the completeness is ~80% at Ks~16 mag, in agreement with the completeness estimation based on the above-described method. (2) We compared our KLF with a KLF obtained from diffraction limited NACO/VLT images (~0.06"angular resolution) of the NSC (Gallego-Cano et al., submitted to A&A) at a projected distance of about 2 pc from Sgr A*. Again, we found a completeness of ~80% at Ks~16 mag, in agreement with our completeness estimation.

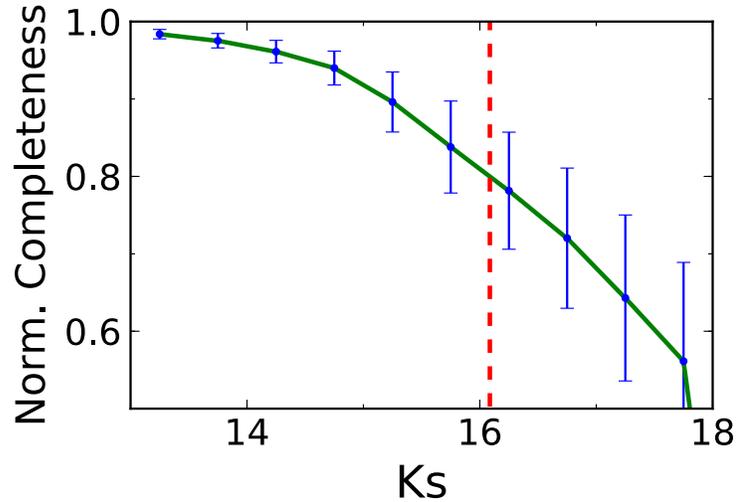

*Fig. 6: Completeness at Ks. The red dashed line shows the 80 % completeness limit.*

Finally, to create the KLF we removed the foreground stars, limited the selection to magnitudes where completeness is > 2/3 (Ks~15.3 mag extinction corrected), and applied the extinction correction with our extinction map.

**V Double Red Clump**

We fitted a simple model of two Gaussians plus an exponential background[10] to the de-reddened KLF (Fig. 7). The two Gaussians correspond to the double RC feature that appears in the CMD. The Double Red Clump is also clearly detected in the region around the Quintuplet cluster by recently published HST observations[33].

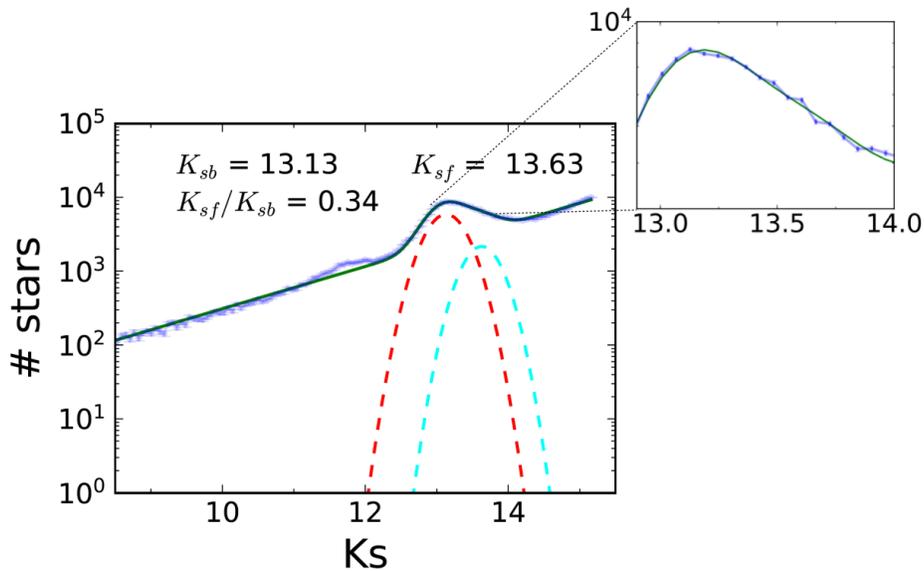

*Fig. 7: KLF obtained with the Ks de-reddened data applying the completeness correction. In green a two Gaussian plus an exponential background fit. Red and cyan dashed lines show the position of the Gaussians. The inset shows the quality of the data in detail, including the error bars and model fit.*

## VI Star Formation History

We fitted the KLF with linear combinations of theoretical models. Tracers that constrain the SFH significantly are the relative changes in magnitude and number of stars between three different remarkable features in the LFs[10]: the asymptotic giant branch bump, the red clump bump and the red giant branch bump (RGBB).

We assumed a Salpeter initial mass function for the models. Initially, we explored the age-metallicity space by using between two and five single age populations and three different metallicities (Z = 0.02, 0.03 and 0.04). We used the same metallicity for all populations in each fit. The best fit was determined by Chi squared minimisation. The bulk of best fits required a metallicity of Z = 0.04, consistent with the metallicity of the innermost bulge, adjacent to the GC region[10]. Although we found that a combination of five ages can already provide us with a satisfactory fit to the measured luminosity function (characterised by a reduced Chi squared value close to 1), a priori, a more complex star SFH cannot be excluded. To analyse the SFH more in detail, we resorted to Monte Carlo (MC) simulations to explore the parameter space spanned by the ages and masses of the different star formation events. Additional fitting parameters were a constant interstellar extinction toward the GC (with differential offsets in extinction having already been accounted for through our use of extinction maps), and a parameter for Gaussian smoothing across bins to absorb magnitude variations due to variable distances and measurement uncertainties. From the measured KLF and assuming Gaussian uncertainties, we created 1000 random KLFs and fitted them with linear combinations of theoretical single age KLFs of ages 13.5, 11, 8, 5, 3, 1, 0.8, 0.6, 0.4, 0.2, 0.1, 0.08, 0.05, 0.03 Gyr (BaSTI[12] models). The oldest isochrone was assumed to be enhanced in alpha elements[10]. We selected the fit with the least Chi squared value for each of the 1000 simulated measurements and thus obtained probability distributions for the ages and masses. Because of degeneracies between the contributions of model populations with similar ages (and thus similar LFs), which can be significant for ages > 8 Gyr, we summed the contributions of the different populations into larger bins for Fig. 2. To obtain the optimum bin widths, we simulated different SFH with uncertainties equivalent to the ones in real data and apply our technique. Thus, we selected the bin width according to the minimum uncertainties for a given bin (Sec. XI for more details).

We addressed potential sources of systematic effects and checked that the derived SFH does not change significantly:

**(1)** Potential contamination by foreground stars. As discussed above, contamination by stars from the Galactic Disk is negligible and contamination from the innermost bar/bulge may be on the order of a few percent. We repeated our analysis by varying the colour-cut used for the exclusion of foreground stars between *H-Ks* = 1.1 and *H-Ks* = 1.5 and did not find any significant effects. As discussed above, there will be stars from the innermost bar present throughout the NSD, but this population is ~12 Gyr old, metal rich, and single age[10] and will thus not affect our conclusions in any significant way.

**(2)** Depending on the precise value of the (*H-Ks*) colour that is used to de-select foreground stars, more or fewer stars from the Galactic bulge may contaminate our nuclear bulge sample. With a stricter criterion (*H-Ks>1.5, upper left panel in Fig. 1*) we can minimise the contribution from inner bulge stars. For this LF, we produced a new extinction map using only RC stars with *H-Ks*>1.5. The SFH resulting from model fitting to the data with this criterion is very similar, but results in a shift of the oldest nuclear bulge stars toward somewhat (by ~2 Gyr) younger ages. This is not surprising given that the innermost bulge (the main contaminant of the nuclear bulge sample) has recently been found to have an age of about 13 Gyr[10].

**(3)** Using MIST isochrones[13,14,15,16,17] (http://waps.cfa.harvard.edu/MIST/) as an alternative to the BaSTI models, we explored the influence of the assumed stellar evolution models on our results. While there are some shifts in the best-fit ages, the overall result is, again, unchanged (Fig. 2). We note, however that the age of the intermediate age stars shifts from 600 Myr to 1 Gyr when using MIST isochrones.

**(4)** We also selected different bin widths for the KLFs, 0.03, 0.06 and 0.1 mag, and confirmed that the results do not change in any significant way.

**(5)** We checked that changing the limit to exclude dark cloud regions does not affect the results.

**(6)** We studied the effect of the completeness limit on the results applying a completeness limit of 70 %. We repeated the analysis using BaSTI models and we did not find any significant variation with respect to the results that we obtained with the 2/3 limit applied previously.

**(7)** We produced different extinction maps for each of the tests involving different selections of stars. In this way, we created extinction maps considering different criteria for the colour cut H-Ks (2), different constrains for excluding the dark clouds regions (5) and a different radius for close stars of 10'' instead of 5''. We did not find any significant change in the results.

**(8)** The majority of stars covered by our sample belong to the red giant branch. This means that their stellar evolution is so fast that assuming different IMFs does not change the results (Fuel consumption theorem[34]). To check it, we repeated the analysis using BaSTI models but considering an IMF with alpha = 1.9. We did not find any significant difference.

**VII Comparison of theoretical models**

We used models from two independent groups to check the possible systematics introduced by the use of any specific theoretical model. We sampled the age parameter according to the differences that appear between the main features present in the LF (AGBB, RC and RGBB) and the relative fraction of stars between them. The LFs change slowly for ages > 5 Gyr, but very rapidly for ages < 2 Gyr which is why we sample young ages more densely. We realised that there are significant differences between the BaSTI and the MIST LFs of certain ages, which will result in age shifts in the inferred SFH. Therefore, we compared the models and selected a slightly different sample of ages for each of them to equally cover the age-space parameter. Fig. 8 depicts a comparison between BaSTI and MIST models for different ages with the same metallicity. As can be seen, at ages < 2 Gyr, the MIST LFs that resemble BaSTI LFs correspond to a stellar populations around twice as old.

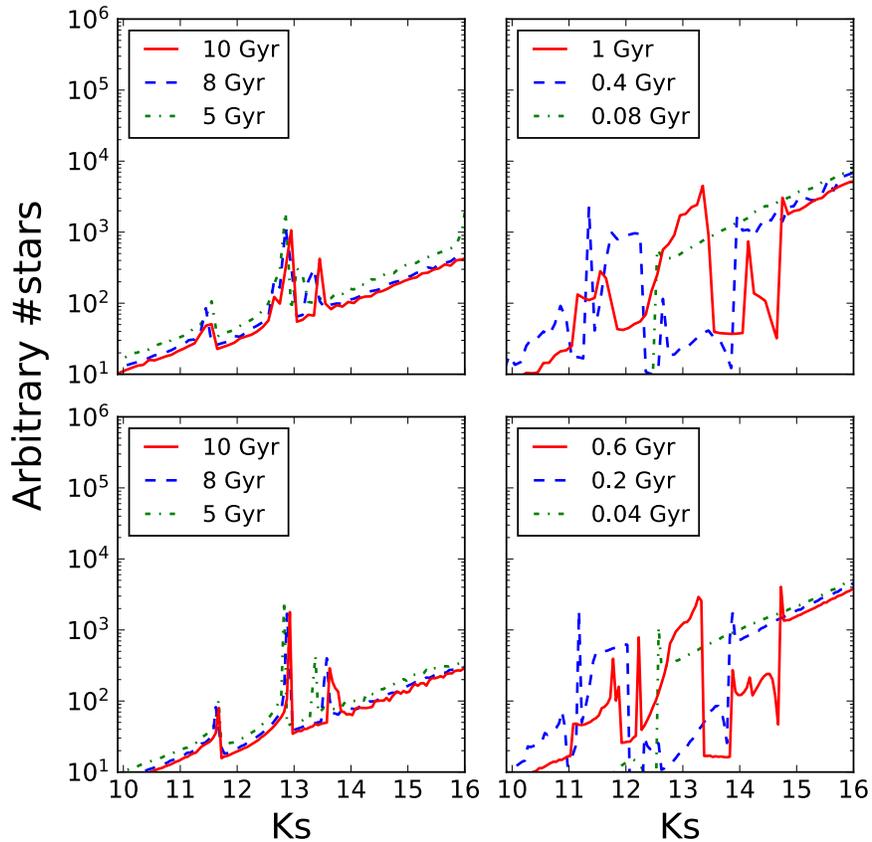

*Fig. 8: Comparison between MIST models (upper panels) and BaSTI models (lower panels) of different ages and twice solar metallicity.*

**VIII Metallicity of the stellar population**

We assumed twice solar metallicity for all theoretical LFs used to derive the SFH. This has to be interpreted as a *mean* metallicity and does not exclude that the underlying population can show a relatively broad range of metallicities. Spectroscopic studies support that the metallicity of stars of both very old and young ages at the GC have a mean (super-) solar metallicity[35,36,37,38,39,40]. Twice solar metallicity was also found in a photometric study of the innermost bulge, 60-90 pc to Galactic north of Sgr A*[10]. Our methodology cannot distinguish between different metallicities at levels more than twice above solar. This is due to the fact that the relative distance between the main features in the KLF for young ages (AGBB, RC and RGBB) and their relative stellar fraction do not change much for high metallicities (Fig. 9).

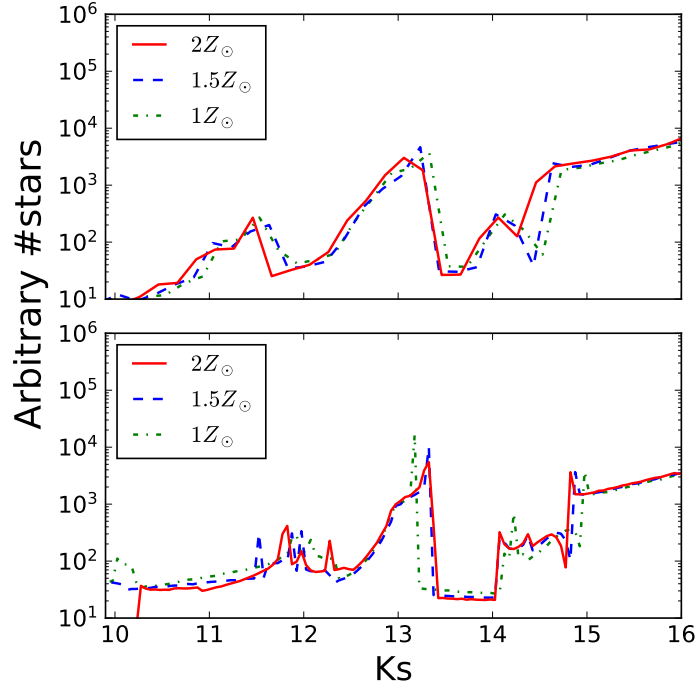

Fig. 9: Theoretical ~1 Gyr KLFs with different metallicities. Upper panel: MIST models. Lower panel: BaSTI models.

We have also checked whether a bulk of solar metallicity stars could explain the observed KLF. However, we found that the residuals are significantly larger than when using twice the solar metallicity and the Chi square for each of the MC samples increases its value by ~30%. Nevertheless, the computed SFH agrees within the uncertainties with the derived assuming twice the solar metallicity. Finally, changing the metallicity of the oldest population in our fits to solar (or sub-solar, Z = 0.01 for BaSTI models) does not change the inferred SFH in any significant way. We concluded that any reasonable variations of the metallicity of the stellar populations do not have any significant impact on the SFH derived by us.

**IX Alpha-element enhancement of the stellar population**

We have assumed the oldest LF in our BaSTI model to be alpha element enhanced ([α/Fe] = +0.4) to account for the inner bulge population[10, 39]. On the other hand, we have used non-alpha element enhanced isochrones to fit the stellar population at the NSD. In this way, we have restricted the parameter space and have decreased the computing time to make it affordable. Nevertheless, we have also carried out several tests to check the effect of an alpha element enhanced stellar population at the NSD. In this way, we performed a fit using 5 models varying their ages to cover the age-space parameter, with twice the solar metallicity and with the possibility of being alpha element enhanced or not. We found that the best solution obtained agrees with the scenario that we found using our 14 models fitting. Namely, we found that the minimum chi-square is obtained for a bulk of old stars (alpha element enhanced in this case), followed by a quiescence period, ended by a intermediate-age star formation burst (non-alpha enhanced) and some young star formation (alpha-enhanced). Moreover, we also tested a simple scenario using all the models with an alpha element enhanced population ($[\alpha/Fe] = +0.4$). We obtained that the chi-square increases its value in more than 200 % with respect to the best fit obtained using a non-alpha enriched population for the NSD.

**X Comparison with previous results**

The SFH that we found disagrees with the so far accepted paradigm of a quasi-continuous star formation with a roughly constant rate at the GC[4]. To cross check our results and methodology we applied our method to the previously used NICMOS data set[4]. We created extinction maps for each field using the same approach described previously. We combined the ~6000 stars detected in all the NICMOS fields and generated an extinction corrected F205W LF, applying a completeness correction and removing foreground stars as explained before. We found that the SFH is compatible with both a constant SFH and the one that we obtain with our data (Fig. 10). This is due to the large uncertainty associated with each LF bin given the low number of stars present in the small regions used. Our GALACTICNUCLEUS data sets includes ~100 times more stars and allowed us to really constrain the uncertainties pointing towards a non-constant SFH.

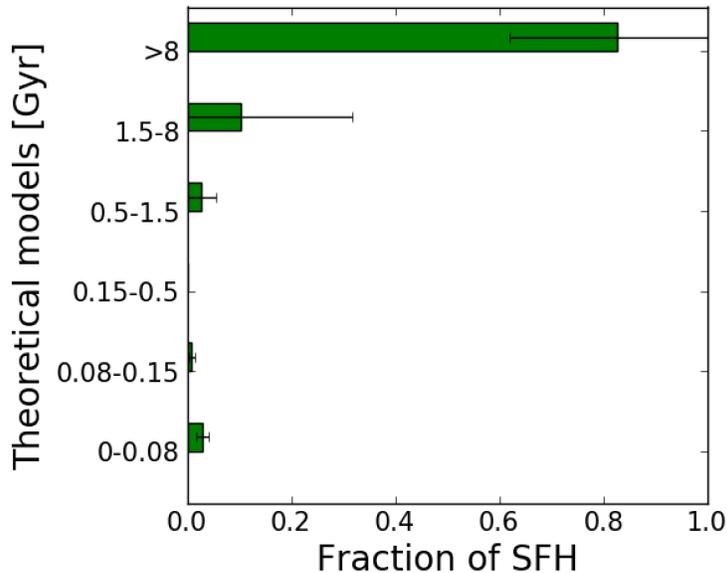

*Fig. 10: SFH obtained for NICMOS data using BaSTI models.*

The previous work[4] also assumed a top-heavy initial mass function (IMF). Thus, we have repeated our analysis using BaSTI models with an IMF of alpha = -1.9 (the equivalent Kroupa/Salpeter value would be alpha = -2.35). We found no significant changes with respect to the results that we have obtained.

On the other hand, we compared out results with the SFH inferred from spectroscopic observations of the Nuclear Star Cluster[41]. They found a non-continuous SFH, with >80 % of the stars formed > 5 Gyr ago, a minimum around 1 Gyr and increased star formation in the past 100 Myr. This is approximately consistent with our findings for the NSD, which provide a higher time resolution than the spectroscopic study. Differences are that in the SFH of the NSD derived here we find that the majority of stars are at least 8 Gyr old and that a negligible fraction of the stars formed between 1.5 and 8 Gyr ago. The spectroscopic study of the NSC does not find the "1 Gyr event" that has left a clear mark in the NSD. There appears to be no secondary RC feature in the NSC. Hence, our results suggest that there are some differences in the SFH of the NSC and NSD, but more detailed spectroscopic and photometric studies are necessary to confirm this.

**XI Tests with artificial SFHs**

To assess the correct performance of our method, we tested it trying to recover artificially created SFHs. We generated three SFHs simulating different scenarios: (1) A non-continuous SFH, similar to the one derived in this work, (2) a continuous constant SFH, and (3) a continuous exponentially decreasing SFH, using BaSTI models. Then, we produced mock KLFs introducing noise according to the real one. Applying our method, we are able to recover the different SFHs within the uncertainties and to distinguish between the different scenarios, in particular between the quasi-continuous and the non-continuous SFHs (Fig. 11). This shows that the small uncertainties of our data allow us to overcome possible degeneracies between similar models.

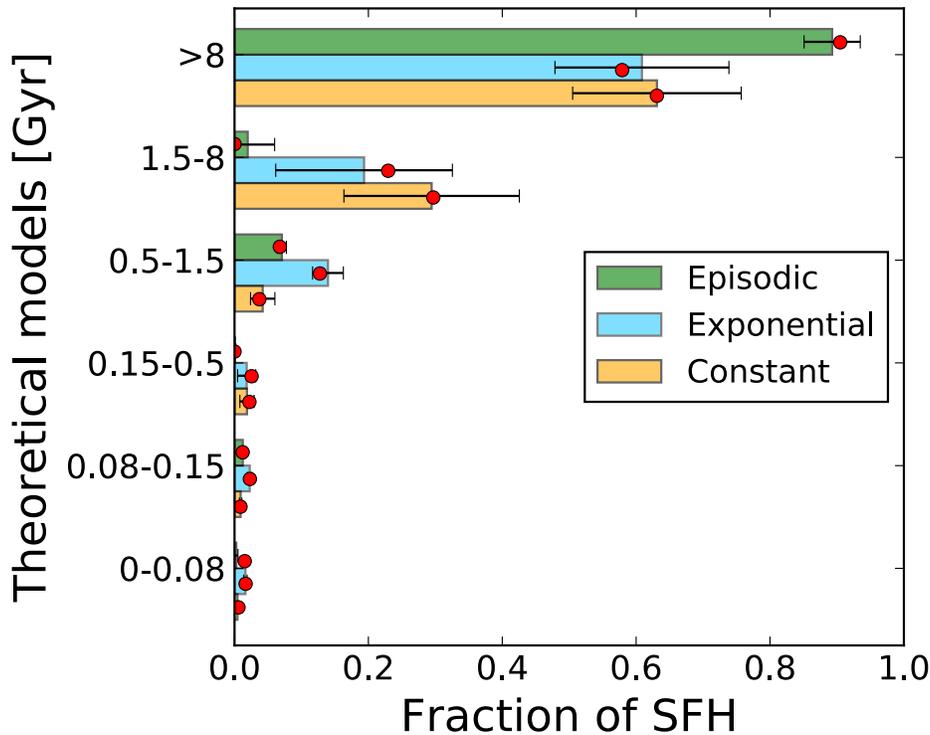

*Fig. 11: Recovery of simulated star formation histories. Red dots: Assumed SFH. Green, blue and orange bars: Recovered star formation.*

**XII Properties of the 1 Gyr event**

To estimate the uncertainty in age of the star formation event at 600 Myr/1 Gyr, we used the probability distributions computed for the MC simulations carried out with BaSTI models (Fig. 12). The bin that includes the star formation event is composed of three models (1 Gyr, 800 Myr and 600 Myr). We found that for the vast majority of MC samples the only component that contributes to the SFH is the 600 Myr one. Since the RC peak is very sensitive to small age changes for young stellar models (Sec. VII) and we have considered for the full fitting procedure models of 800 and 400 Myr besides of the most probable of 600 Myr, we estimate that the uncertainty in age cannot be larger than 100 Myr for the detected star formation event.

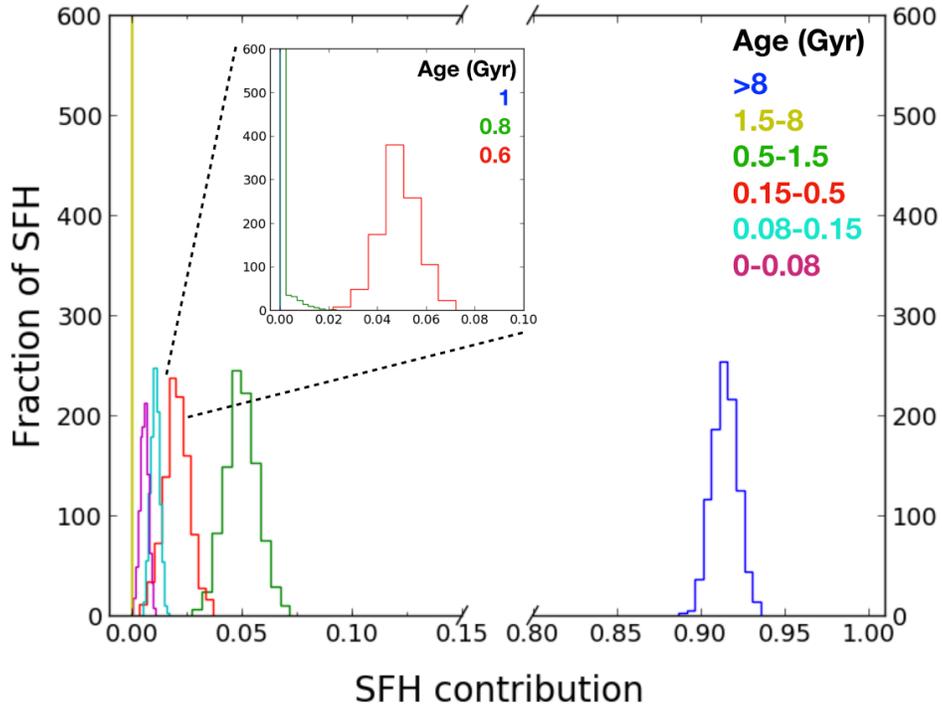

*Fig. 12: Probability distribution for the MC simulations carried out using BaSTI models. The legend indicates the contribution of each age bin. The inset shows all the models contributing to the bin where we detect the star formation event at ~1Gyr.*

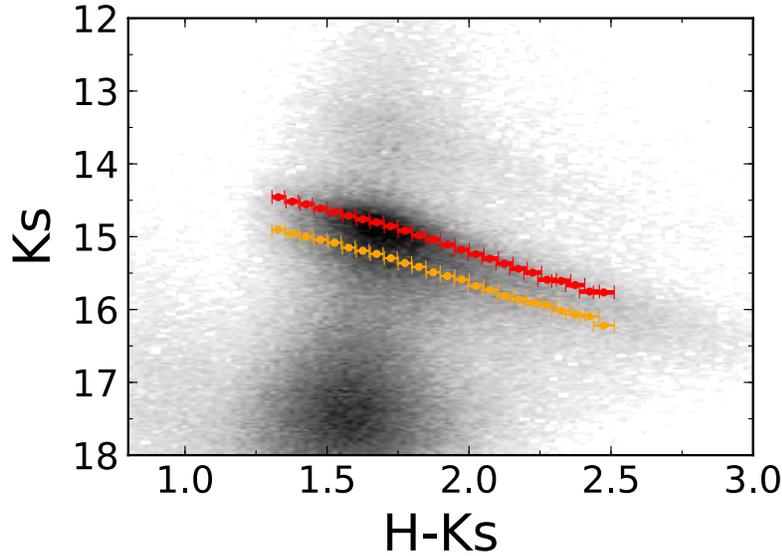

*Fig. 13: CMD Ks versus H − Ks. Red and orange points and uncertainty bars depict the peaks of the best double-Gaussian fit found for each bin of H – Ks.*

A direct confirmation of the detected intermediate age star formation event is the presence of a secondary feature in the RC in the CMD that is indicated by the red arrows in Fig. 1. To check that the secondary feature is real, we defined a region that includes both features containing stars with $H - Ks \in$ [1.3, 2.5] and divided it into $H - Ks$ bins of 0.05 mag. For each of them, we fitted the $Ks$ distribution with a single and a double Gaussian model applying the expectation maximisation algorithm implemented in the SCIKIT-LEARN python object GaussianMixture. Using the *Bayesian Information Criterion* and the *Akaike Information Criterion*, we confirmed that a combination of two Gaussians provides a better fit for the data. Both features run parallel to the reddening vector with slopes of 1.19 ± 0.02 ± 0.01 and 1.15 ± 0.02 ± 0.01 for the first and the second feature, respectively, as is shown in Fig. 13. The errors refer to the statistical and the systematic and were computed using a jackknife resampling method, different bin widths and lower cuts for the RC region employed. The distance between features is $\Delta Ks$ = 0.44±0.01. We also estimated the extinction to each feature assuming an extinction index of 2.30±0.08 and a grid of reddened intrinsic colour $(H-Ks)_0$ in steps of 0.01 mag. We minimised the difference between data and the model and obtained very similar extinction values of 1.92±0.14 mag and 1.94±0.16 mag, for the bright and the faint feature, respectively. These similar values allowed us to exclude the alternative scenario of a stellar population beyond the GC to explain the secondary clump. Finally, we also tried to fit a single theoretical LF to the data to exclude the RGBB[10] as the explanation of the secondary feature. We did not get any satisfactory fit. Thus, although RGBB stars must necessarily be present (and are taken into account in the theoretical models used here), it is necessary to consider a secondary RC to fully explain the detected feature.

**XIII CMD analysis**

The lower completeness in *H* band due to the lower sensitivity impedes to perform an analysis of its luminosity function as we have shown for the *K* band. Instead, to assess the obtained SFH, we also simulated a CMD *Ks* versus *H-Ks* using BaSTI models. We added random noise to the models in *H* and *Ks* bands according to the uncertainty of the data, and took into account the reddening using the computed extinction of the RC stars in our sample. We based our analysis exclusively on the relative distance between the RC features to avoid problems related to completeness. From our LF fit we found that the main contributions to the double RC feature arise from the oldest stars (≳80% of the mass is ≳8 Gyr) and from the star formation event at ~1 Gyr. Therefore, we simulated several synthetic CMDs to check this scenario:

(1) We simulated just an old population of 8 Gyr to study the contamination by the RGBB. We analysed the RC features using the SCIKIT-LEARN python object GaussianMixture as explained in the previous section. We detected a secondary clump that corresponds to the RGBB at $\Delta$Ks = 0.62±0.02 mag. This scenario is not compatible with the real data since the relative distance between RC and RGBB is 0.18 ± 0.02 mag larger than in the data, where the uncertainties from real data and the simulation have been propagated quadratically. We concluded that there is some contamination from the RGBB but it is not enough to explain the double RC detected with our analysis.
(2) We generated a SFH similar to the one obtained using stellar models of 8 Gyr, 600 Myr and 40 Myr. We obtained that there is a secondary RC as the one observed in the real data. We computed the distance between features and obtained $\Delta$Ks = 0.46±0.03 in good agreement with the result obtained for real data ($\Delta$Ks = 0.44±0.01).
(3) We simulated a continuous SFH with a constant star formation rate. We used 5 different models from 0.1 to 10.1 Gyr equally spaced in age and combined them. We repeated the previous analysis and found a two feature structure with $\Delta$Ks = 0.61±0.03 mag, where the secondary clump is due to the RGBB as in model (1) and incompatible with the data.
(4) Finally, we created an exponential star formation history using BaSTI models of 10 Gyr, 8 Gyr, 5 Gyr, 3 Gyr, 1 Gyr, 600 Myr, 200 Myr and 30 Myr. We found a secondary clump due mainly to the 600 Myr models. We found $\Delta$Ks = 0.36±0.04, which is smaller than what we measure in the actual data. Moreover, the AGBB is much more prominent in this simulation given the fraction of young stars (Fig. 14d)

Fig. 14 shows the original data and the simulations obtained for (2), (3) and (4). We conclude that only the SFH simulated using the parameters obtained with the KLF analysis is able to reproduce the observed data.

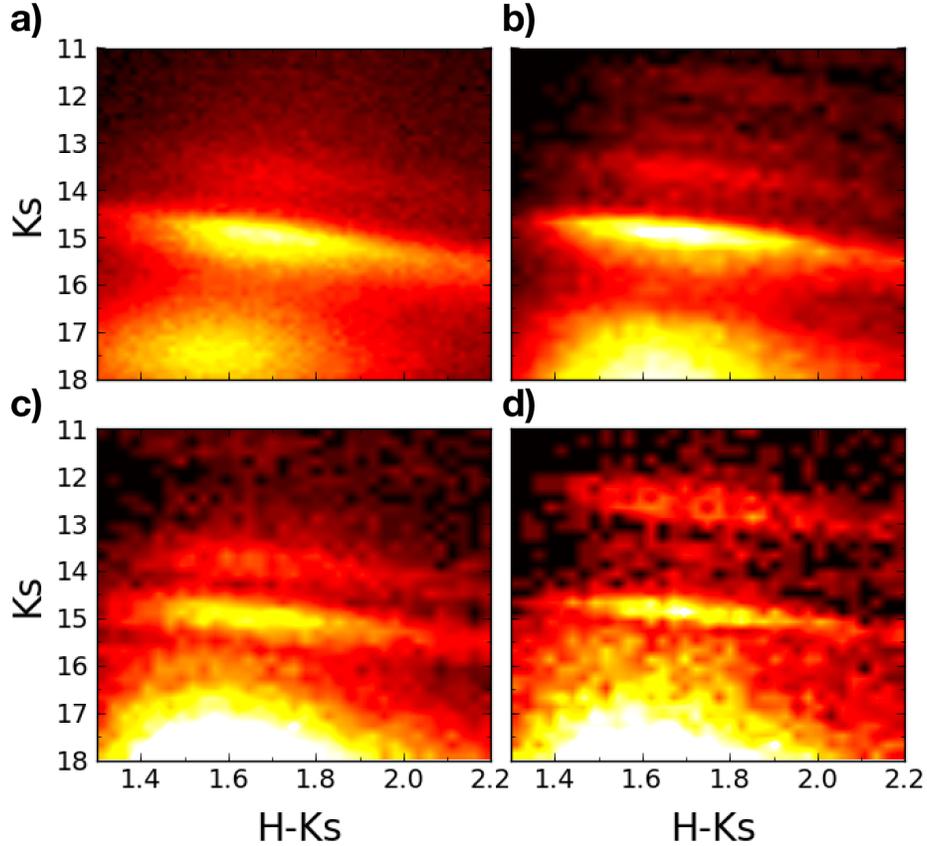

*Fig 14. Comparison between CMDs produced using real data (a) and several simulations using BaSTI models: (b) Best-fit solution obtained from the KLF. (c) Continuous SFH. (d) Exponential SFH.*

**XIV Low–extinction regions analysis**

To assess the obtained results, we repeated the analysis of the luminosity function selecting only low-extinction regions. Since the J band is more prone to extinction given its shorter wavelength, we used the J-band density map approach, described in Sect. II, to select the low-extinction regions. We masked regions with a J-band density below 25 % of the maximum density detected. The selected region comprises ~10 % of the total area covered and is well distributed across the field. We created extinction maps using the selected stars and the technique described in Sect. III. We obtained a KLF removing the foreground population, correcting for differential extinction, and applying the completeness solution, as explained previously. We also created a HLF using the same technique. Comparing both LFs, we checked that the HLF completeness is significantly affected by sensitivity at H>14 mag. We estimated that our completeness correction (that only considers incompleteness due to crowding), overestimates the completeness ~5 % for H~14 mag and ~30 % for H~15 mag. Thus, the faint end of the RC and the RGBB are not adequately covered by the HLF. Given the high incompleteness at H>14 mag, we decided to restrict the faint limit of the HLF to H=14.1 mag and apply a correction of 5 % on the completeness solution and its uncertainties accordingly with the previous estimation. We also repeated the analysis assuming a correction of ~10 % without any significant difference on the results. We applied our full-fitting LF method (Sect. VI) assuming BaSTI models for both, HLF and KLF. Figure 15 shows the obtained results. The results agree with a mainly old population with a quiescence period ~5 Gyr followed by a significant star formation burst ~1Gyr. We believe that the stronger contribution of young stars for the SFH inferred from the HLF is a degeneracy, because bright stars can be either young stars or old giants, and the old population is less well constrained in the HLF because it is shallower than the KLF.

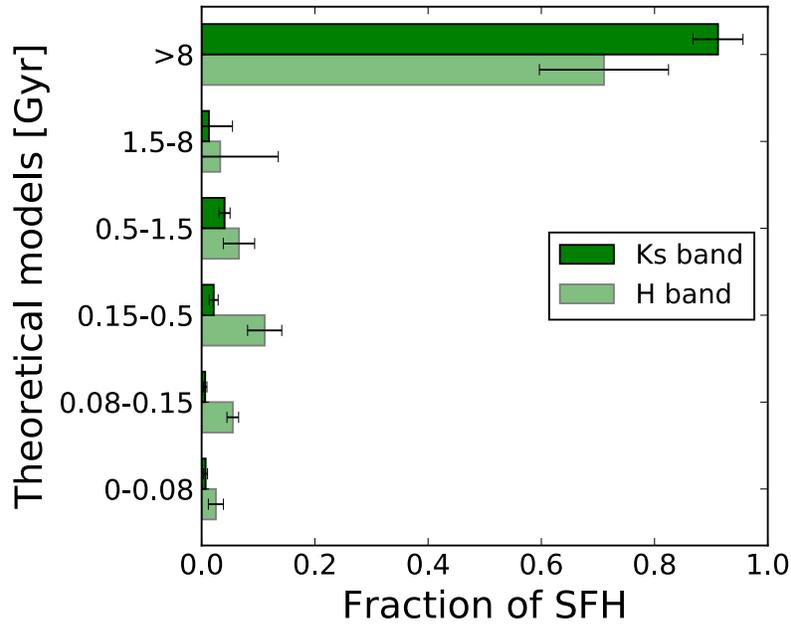

Fig 15. Star Formation History of the low-extinction regions of the NSD using BaSTI models for Ks and H bands.

## XV Mass Estimation

We calculated the total mass of the NSD using the results obtained when fitting the 1000 MC samples with MIST models. For our region, we obtain $(6.5\pm0.4)\times 10^7$ $M_\odot$. The NSD was previously estimated to contain $(8\pm2)\times 10^8$ $M_\odot$ within R = 120 pc of Sgr A*, based on its near-infrared emission[1]. To cross check this number, we use an extinction and PAH-emission corrected Spitzer 4.5 μm image of the GC[42] and integrate the light inside an ellipse centred on Sgr A* with 120 pc radius along the Galactic plane and a minor-to-major axis ratio of 0.36[42], after previous subtraction of a model for the bulge emission (Gallego-Cano et al., submitted to A&A). We then scale with a mass-to-luminosity ratio of 0.6±0.2[43] to obtain a total stellar mass of $(8\pm4)\times 10^8$ $M_\odot$, consistent with the one derived from the near-infrared measurements. After scaling the mass derived from the KLF fits of the region studied here to the nuclear bulge inside of 120 pc, we obtain a consistent value of $(9.8\pm0.6)\times 10^8$ $M_\odot$, which indicates that our basic method and assumptions, in particular about the initial mass function, appear to be adequate.

We computed the total mass of supernovae assuming a standard Salpeter IMF and propagating the uncertainties obtained when considering the total stellar mass formed ~1 Gyr ago (obtained by means of MIST model fitting).